\documentclass[letterpaper,twocolumn,10pt]{article}
\usepackage{usenix2019_v3}
\usepackage[none]{hyphenat}
\usepackage{cite}
\usepackage{amsmath,amssymb,amsfonts}
\usepackage{algorithmic}
\usepackage{graphicx}
\usepackage{textcomp} 
\usepackage{xcolor}
\usepackage{xspace}
\usepackage{framed}
\usepackage{titlesec}

\def\BibTeX{{\rm B\kern-.05em{\sc i\kern-.025em b}\kern-.08em
    T\kern-.1667em\lower.7ex\hbox{E}\kern-.125emX}}
    
\usepackage{booktabs}
\usepackage{subcaption}
\usepackage{xspace}
\usepackage{color}
\usepackage{tabu}
\usepackage{hyperref}
\usepackage{longtable,tabularx}
\usepackage{ragged2e}
\usepackage{float}

\usepackage{url}
\makeatletter
\g@addto@macro{\UrlBreaks}{\UrlOrds}
\makeatother

\usepackage{tikz}
\usetikzlibrary{decorations.pathreplacing}
\usetikzlibrary{fadings}
\usepackage[normalem]{ulem}
\usetikzlibrary{positioning,fit,calc,backgrounds}
\tikzset{block/.style={draw,thick,text width=2cm,minimum height=1cm,align=center},
         line/.style={-latex}
}
\makeatletter
\def\ruwave{\bgroup \markoverwith{\lower3.5\p@\hbox{\textcolor{red}{\sixly \char58}}}\ULon}
\font\sixly=lasy6
\makeatother

\newlength{\bibitemsep}
\newlength{\bibparskip}
\let\oldthebibliography\thebibliography
\renewcommand\thebibliography[1]{
  \oldthebibliography{#1}
  \setlength{\parskip}{\bibitemsep}
  \setlength{\itemsep}{\bibparskip}
}

\newcommand{\point}[1]{\smallskip\par\noindent\textbf{#1}:}

\newcommand{\etal}{et~al.\xspace}

\newcommand{\empirical}[1]{#1}
\newcommand{\emperical}[1]{#1}

\titlespacing\section{0pt}{10pt plus 4pt minus 2pt}{2pt plus 2pt minus 2pt}
\titlespacing\subsection{0pt}{10pt plus 4pt minus 2pt}{2pt plus 2pt minus 2pt}
\titlespacing\subsubsection{0pt}{10pt plus 4pt minus 2pt}{2pt plus 2pt minus 2pt}

\begin{document}
\sloppy

\title{\tool: Making In-Browser Perceptual Ad Blocking \\Practical with Deep Learning}
\author{Zain ul abi Din\footnotemark{} \\UC Davis \and 
Panagiotis Tigas\footnotemark{} \\ University of Oxford \and 
Samuel T. King \\ UC Davis \\ Bouncer Technologies \and 
Benjamin Livshits \\ Brave Software \\ Imperial College London}



\newcommand{\tr}[1]{#1} 
\newcommand{\notr}[1]{} 
\newcommand{\tool}{\textsc{Percival}\xspace}

\newcommand{\easylistaccuracy}{\empirical{96.76\%}}
\newcommand{\facebookaccuracy}{\emperical{95.25\%}}
\newcommand{\portuguesseaccuracy}{\emperical{97.16\%}}
\newcommand{\russianaccuracy}{\emperical{94.90\%}}
\newcommand{\performancerel}{\emperical{4.55\%}}
\newcommand{\performanceabs}{\emperical{178.23ms\xspace}}
\newcommand{\modelsize}{\emperical{1.76MB}}

\maketitle
\pagestyle{plain}

\begin{abstract}

In this paper we present \tool, a browser-embedded, lightweight, deep learning-powered ad blocker. \tool embeds itself within the browser's image rendering pipeline, which makes it possible to intercept every image obtained during page execution and to perform image classification based blocking to flag potential ads.

Our implementation inside both Chromium and Brave browsers shows only a minor rendering performance overhead of~\performancerel, for Chromium, demonstrating the feasibility of deploying traditionally heavy models (i.e. deep neural networks) inside the critical path of the rendering engine of a browser. We show that our image-based ad blocker can replicate EasyList rules with an accuracy of~\easylistaccuracy. 
Additionally, \tool does surprisingly well on ads in languages other than English and also performs well on blocking first-party Facebook ads, which have presented issues for other ad blockers.
\tool proves that image-based perceptual ad blocking is an attractive complement to today's dominant approach of block lists.

\end{abstract}
\footnotetext{Employed by Brave software when part of this work took place.}

\section{Introduction} 
\label{sec:introduction}
Web advertising provides the financial incentives necessary to support most of the free content online, but it comes at a security and privacy cost. To make advertising effective, ad networks or publishers track user browsing behavior across multiple sites to generate elaborate user profiles for targeted advertising. 

Users find that ads are intrusive~\cite{Pujol_annoyedusers:} and cause disruptive browsing experience~\cite{browsingexperience,hate-ads}.
Also, studies have shown that advertisements impose privacy and performance costs to users, and carry the potential to be a malware delivery vector~\cite{Li:2012:KYE:2382196.2382267,xing2015understanding, ads-annoy,Englehardt:2016:OTM:2976749.2978313,Garimella2017,exposingthehiddenweb}. 

Ad blocking is a software capability for filtering out unwanted advertisements to improve user experience, performance, security, and privacy. At present, ad blockers either run directly in the browser ~\cite{mozilla,brave} or as  browser extensions~\cite{adblockplus}. 

Current ad blocking solutions filter undesired content based on ``handcrafted'' filter lists such as EasyList~\cite{vastel2018filters}, which contain rules matching ad-carrying URLs and DOM elements. Most widely-used ad blockers, such as uBlock Origin~\cite{ublock} and Adblock Plus~\cite{adblockplus} use these block lists for content blocking.
While useful, these approaches fail against adversaries who can change the ad-serving domain or obfuscate the web page code and metadata. 

In an attempt to find a more flexible solution, researchers have proposed alternative approaches to ad blocking. One such approach is called Perceptual ad blocking, which relies on ``visual cues'' frequently associated with ads like the AdChoices logo or a sponsored content link. 
Storey \etal~\cite{Storey} built the first perceptual ad blocker that uses traditional computer vision techniques to detect ad-identifiers. Recently, Adblock Plus developers built filters into their ad blocker~\cite{image-search} to match images against a fixed template in order to detect ad labels. Due to variations in ad-disclosures, AdChoices logo and other ad-identifiers, it is unlikely that traditional computer vision techniques are sufficient and generalizable to the range of ads one is likely to see in the wild.

\begin{figure}[t]
    \centering
    \includegraphics[width=1.0\columnwidth]{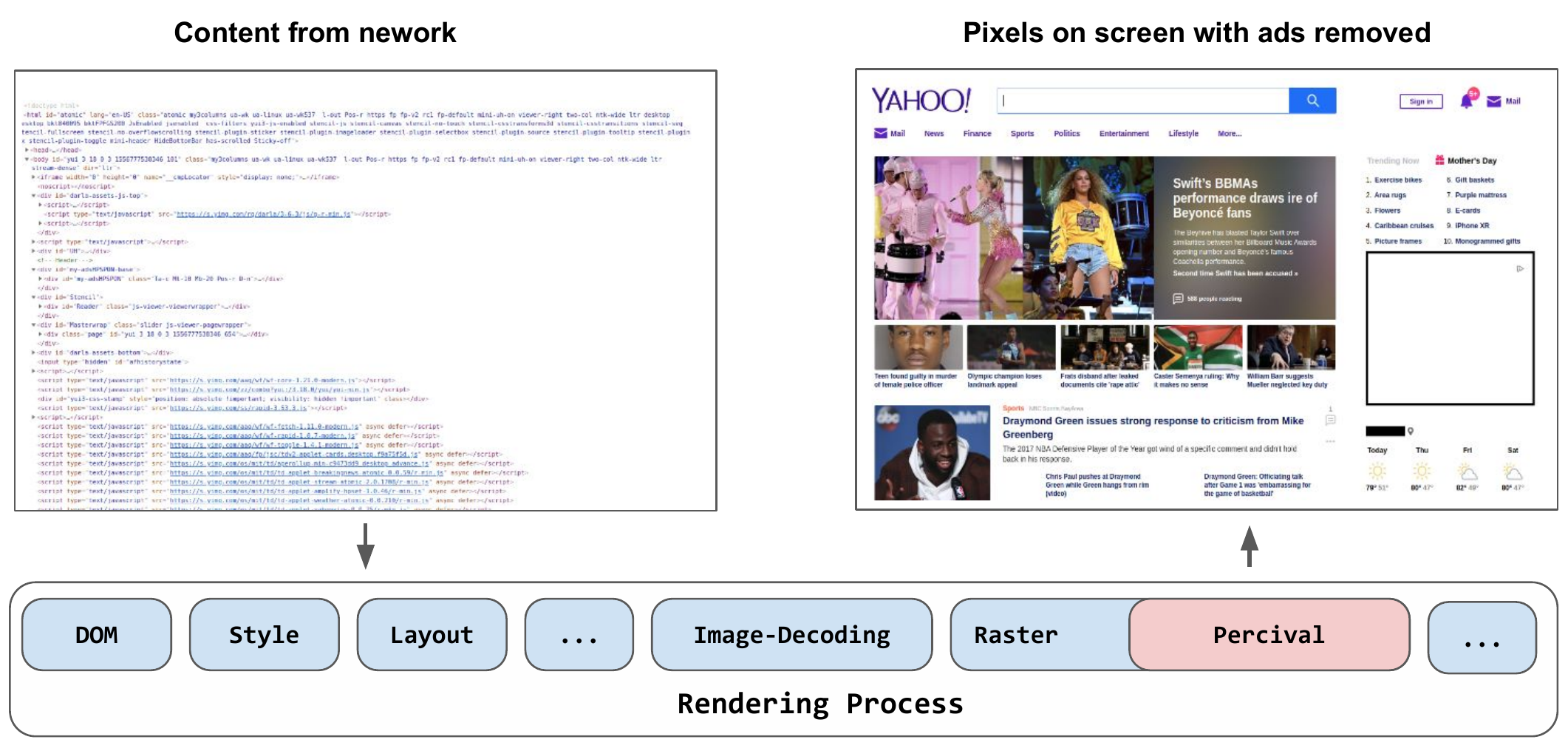}
    \caption{Overall architecture of \tool. \tool is positioned in the renderer process-which is responsible for creating rasterized pixels from HTML, CSS, JavaScript. As the renderer process creates the DOM and decodes and rasterizes all image frames, these are first passed through \tool. \tool blocks the frames that are classified as ads. The corresponding output with ads removed is shown above (right).}
    \label{fig:architecture}
\end{figure}

A natural extension to traditional vision-based blocking techniques is deep learning. Adblock Plus recently proposed SENTINEL~\cite{Sentinel2018} that detects ads in web pages using deep learning. SENTINEL's deep learning model takes as input the screenshot of the rendered webpage to detect ads. However, this technology is still in development, and in its existing form, its machine learning model is prohibitively large for practical deployment.

To this end, we present \tool, a native, deep learning-powered \emph{perceptual ad blocker}, which is built into the browser image rendering pipeline. \tool intercepts every image obtained during the execution sequence of a page and blocks images that it classifies as ads.
\tool is small (half the average webpage size~\cite{page-size}) and fast, and we deploy it online within two commercial browsers to do real-time ad detection and blocking. 

\tool can be run \emph{in addition} to an existing ad blocker, as a last-step measure to block whatever slips through its filters. However, \tool may also be deployed \emph{outside} the browser, for example, as part of a crawler, whose job is to construct comprehensive block lists to supplement EasyList.

\subsection{Contributions}
This paper makes the following contributions:
\begin{itemize}\itemsep=-1pt
\item {\bf Perceptual ad blocking in Chromium-based browsers.}
We deploy \tool in two Chromium-based browsers: Chrome and Brave.
We demonstrate two deployment scenarios; first, we block ads synchronously as we render the page, with a modest performance overhead. Second, we classify images \emph{asynchronously} and memoize the results, thus speeding up the classification process\footnote{We make the source code, pre-trained models and data available for other researchers at \url{https://github.com/dxaen/percival}}.

\item {\bf Lightweight and accurate deep learning models.}
We show that ad blocking can be done effectively using highly-optimized deep neural network-based models for image processing. Previous studies suggest that models over~5MB in size become hard to deploy on mobile devices~\cite{play_store_apk_size}; because of our focus on low-latency detection, we create a compressed in-browser model that occupies  1.76MB\footnote{Our in-browser model is 3.2MB due to a less efficient serialization format. Still, the weights are identical to our 1.76MB model} on disk, which is smaller by factor of~\empirical{150} compared to other models of this kind~\cite{sentinel}, while maintaining similar accuracy results.


\item {\bf Accuracy and performance overhead measurements.}
We show that our perceptual ad blocking model can replicate EasyList rules with the accuracy of~\easylistaccuracy, making \tool a viable and complementary ad blocking layer. 
Our implementation within Chromium shows an average overhead  of~\performanceabs for page rendering.  This overhead shows the feasibility of deploying deep neural networks inside the critical path of the rendering engine of the browser. 

\item {\bf First-party ad blocking.}
While the focus of traditional ad blocking is primarily on third-party ad blocking, we show that \tool blocks first-party ads as well, such as those found on Facebook. Specifically, our experiments show that \tool blocks ads on Facebook (often referred to as  ``sponsored content'') with a~\empirical{92\%} accuracy, 
with precision and recall of~\empirical{78.4\%} and~\empirical{70.0\%}.

\item {\bf Language-agnostic blocking.} 
We demonstrate that our model in \tool blocks images that are in languages we did not train our model on. We evaluate our trained model on Arabic, Chinese, Korean, French and Spanish datasets. Our model achieves an accuracy of~\empirical{81.3\%} on Arabic,~\empirical{95.1\%} on Spanish, and~\empirical{93.9\%} on French datasets, with moderately high precision and recall. However, results from Chinese and Korean ads are less accurate.
\end{itemize}



\section{Motivation}
Online advertising has been a long standing concern for user privacy, security and overall web experience. While web advertising makes it easier and more economic for businesses to reach a wider audience, bad actors have exploited this channel to engage in malicious activities. Attackers use ad-distribution channels to hijack compromised web pages in order to trick users into downloading malware~\cite{Li:2012:KYE:2382196.2382267}. This is known as malicious advertising.

Mobile users are also becoming targets of malicious advertising~\cite{mobileads}. Mobile applications contain code embedded from the ad networks, which provides the interface for the ad networks to serve ads. This capability has been abused by attackers where the landing page of the advertisements coming from ad networks link to malicious content. Moreover, intrusive advertisements significantly affect the user experience on mobile phones due to limited screen size~\cite{Goldstein2013TheCO}. Mobile ads also drain significant energy and network data~\cite{breakingforcommercials}.

Web advertising also has severe privacy implications for users. Advertisers use third party web-tracking by embedding code in the websites the users visit, to identify the same users again in a different domain, creating a more global view of the user browsing behavior~\cite{internetjones}. Private user information is collected, stored and sold to other third party advertisers. These elaborate user profiles can be used to infer sensitive information about the users like medical history or political views~\cite{blockme, betrayed}. Communication with these third party services is unencrypted, which can be exploited by attackers.

The security and privacy concerns surrounding web advertising has motivated research in ad blocking tools from both academia~\cite{adgraph, Bhagavatula2014,Gugelmann2015,mltracking, shuba2020nomoats} and industry notably Adblock Plus~\cite{adblockplus}, Ghostery~\cite{ghostery}, Brave~\cite{brave},  Mozilla~\cite{Kontaxis2015TrackingPI}, Opera~\cite{operanative} and Apple~\cite{appleadblock}. 
Ad blocking serves to improve web security, privacy, usability, and performance. As of February~2017,~615 million devices had ad blockers installed~\cite{adblockreport}
However, recently Google Chrome~\cite{chromeblock} and Safari~\cite{safariblock} proposed changes in the API exposed to extensions, with the potential to block extension based ad-blockers. This motivates the need for native ad blockers like Brave~\cite{brave}, Opera~\cite{operanative}, AdGraph~\cite{adgraph} and even \tool.

\section{\tool Overview}
\label{sec:overview}

This paper presents \tool, a new system for blocking ads. Our primary goal is to build a system that blocks ad images that might be allowed by current detection techniques, while remaining small and efficient enough to run in a mobile browser.
Figure~\ref{fig:architecture} shows how \tool blocks rendering of ads. First, \tool runs in the browser image rendering pipeline. By running in the image rendering pipeline, we can ensure that we inspect all images before the browser shows them to the user. Second, \tool uses a deep convolutional neural network~(CNN) for detecting ad images. Using CNNs enables \tool to detect a wide range of ad images, even if they are in a language that \tool was not trained on.

This section discusses \tool's architecture overview, possible alternative implementations and detection model. Section~\ref{sec:design} discusses the detailed design and implementation for our browser modifications and our detection model.

\subsection{\tool's Architecture Overview}
\label{subsec:browser_architecture}

\tool’s detection module runs in the browser’s image decoding pipeline after the browser has decoded the image into pixels, but before it displays these pixels to the user. Running \tool after the browser has decoded an image takes advantage of the browser’s mature, efficient, and extensive image decoding logic, while still running at a choke point before the browser displays the decoded pixels. Simply put, if a user sees an image, it goes through this pipeline first.

More concretely, as shown in Figure~\ref{fig:architecture} \tool runs in the renderer process of the browser engine. The renderer process on receiving the content of the web page proceeds to create the intermediate data structures to represent the web page. These intermediate representations include the DOM-which encodes the hierarchical structure of the web page, the layout-tree, which consists of the layout information of all the elements of the web page, and the display list, which includes commands to draw the elements on the screen. If an element has an image contained within it, it needs to go through the \emph{Image Decoding Step} before it can be rasterized. We run \tool after the \emph{Image Decoding Step} during the \emph{raster} phase which helps run \tool in parallel for multiple images at a time. Images that are classified as ads are blocked from rendering. The web page with ads removed is shown in Figure~\ref{fig:architecture} (right). We present the detailed design and implementation in Section~\ref{sec:design}

\subsection{Alternative Possible Implementations and Advantages of \tool}
One alternative to running \tool directly in the browser could have been to run \tool in the browser’s JavaScript layer via an extension. However, this would require scanning the DOM to find image elements, waiting for them to finish loading, and then screenshotting the pixels to run the detection model.
The advantage of a JavaScript-based system is that it works within current browser extensibility mechanisms, but recent work has shown how attackers can evade this style of detection~\cite{Tramer2018}. 

Ad blockers that inspect web pages based on the DOM such as Ad Highlighter~\cite{Storey} are prone to DOM obfuscation attacks. They assume that the elements of the DOM strictly correspond to their visual representation. For instance, an ad blocker that retrieves all \texttt{img} tags and classifies the content contained in these elements does not consider the case, where a rendered image is a result of several CSS or JavaScript transformations and not the source contained in the tag. These ad blockers are also prone to resource exhaustion attacks where the publisher injects a lot of dummy elements in the DOM to overwhelm the ad blocker.

\begin{figure}[tb]
\centering
    \includegraphics[width=0.6\columnwidth]{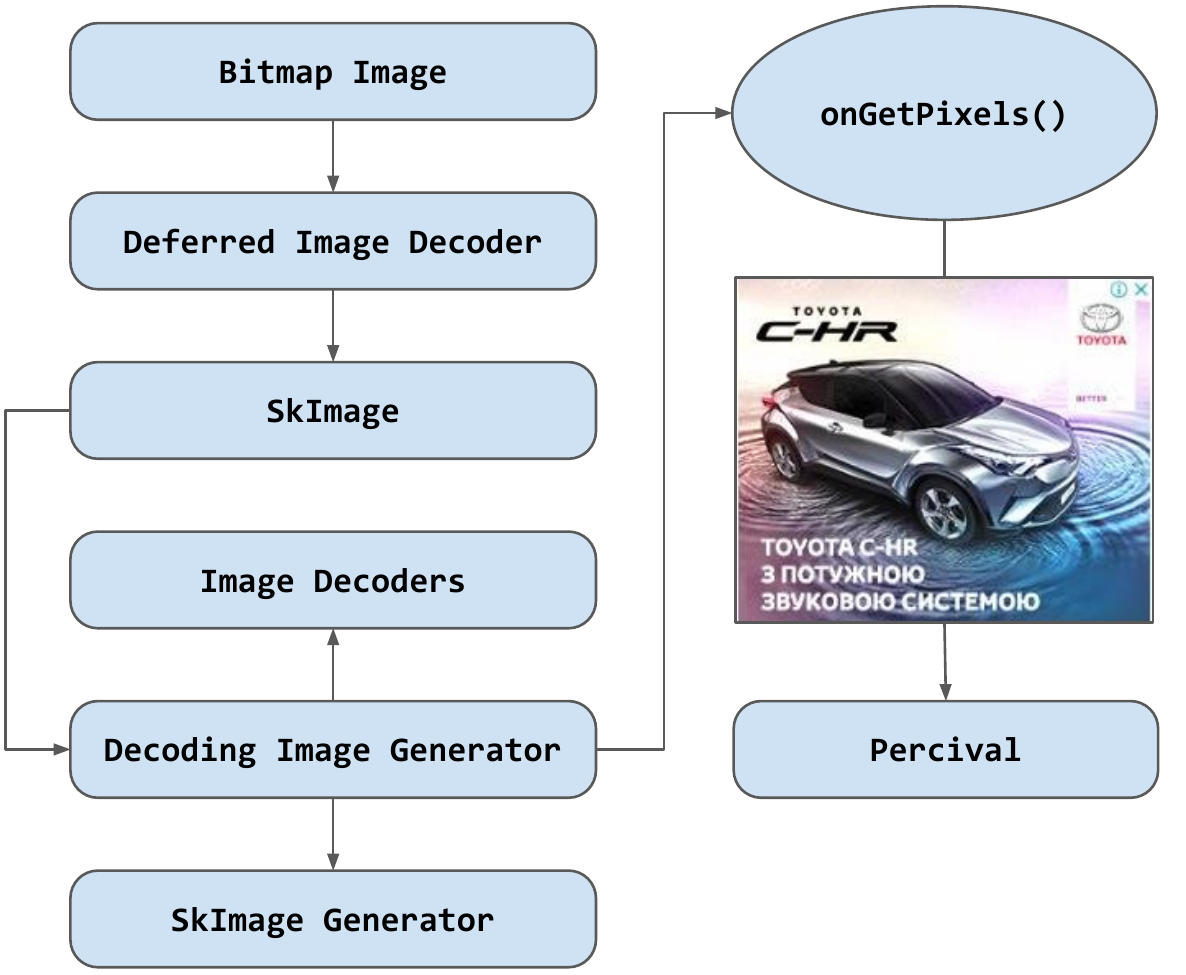}
    \caption{\tool in the image decoding pipeline. SkImage Generator allocates a bitmap
    and calls the \texttt{onGetPixels()} of \texttt{DecodingImageGenerator} to populate the bitmap. This bitmap is then passed to the network for classification and cleared if it contains an ad.}
    \label{fig:inst}
\end{figure}

Additionally, a native implementation is much faster than a browser extension implementation with the added benefit of having access to the unmodified image buffers. 
\subsection{Detection Model}
\label{subsec:detection_model}
\tool runs a detection model on every image loaded in the document's main frame, a sub-document such as an \texttt{iframe}, as well as images loaded in JavaScript to determine if the image is an ad.

Although running directly within the browser provides \tool with more control over the image rendering process, it introduces a challenge: how to run the model efficiently in a browser. Our goal is to run \tool in browsers that run on laptops or even mobile phones.  This requires that the model be small to be practical~\cite{play_store_apk_size}.  This design also requires that the model run directly in the image rendering pipeline, so overhead remain low.  Any overhead adds latency to rendering for all images it inspects.

In \tool, we use the SqueezeNet~\cite{Iandola2016} CNN as the starting point for our detection model. 
We modify the basic SqueezeNet network to be optimized for ad blocking by removing less important layers.  This results in a model size that is less than~2~MB and detects ad images in~\empirical{11~ms}.

A second challenge in using small CNNs is how to provide enough training data. In general, smaller CNNs can have suitable performance but require more training data. Gathering ad images is non-trivial for a number of reasons, the number of ad and non-ad images on most web pages is largely skewed in favor of non-ad images. 
Most ads are programmatically inserted into the document through \texttt{iframe}s or JavaScript, and so simple crawling methods that work only on the initial HTML of the document will miss most of the ad images.

To crawl ad images, other researchers~\cite{Tramer2018, sentinel} propose screenshotting iframes or JavaScript elements. This data collection method leads to problems with synchronizing the timing of the screenshot and when the element loads. Many screenshots end up with whites-space instead of the image content. Also, this method only makes sense if the input to the classifier is the rendered content of the web page.

To address these concerns and to provide ample training data, we design and implement a custom crawler in Blink that handles dynamically-updated data and eliminates the race condition between the browser displaying the content and the screenshot we use to capture the image data. Our custom-crawler downloads and labels ad and non-ad images directly from the rendering pipeline.

\section{Design and Implementation of \tool}
\label{sec:design}

This section covers the design and implementation of the browser portion of \tool. We first cover the high-level design principles that guide our design, then we discuss rendering and image handling in Blink, the rendering engine of Google Chrome and the Brave browser. Finally, we describe our end-to-end implementation within Blink.

\subsection{Design Goals}
We have two main goals in our design of \tool: 
\point{Run \tool at a choke point} Advertisers can serve ad images in different formats, such as JPG, PNG, or GIF. Depending on the format of the image, an encoded frame can traverse different paths in the rendering pipeline. Also, a wide range of web constructs can cause the browser to load images, including HTML image tags, JavaScript image objects, HTML Canvas elements, or CSS background attributes. Our goal is to find a single point in the browser to run \tool, such that it inspects all images, operates on pixels instead of encoded images, but does so before the user sees the pixels on the screen, enabling \tool to block ad image cleanly. \textbf{Note:} If individual pixels are drawn programmatically on canvas, \tool will not block it from rendering.

In Blink, the raster task within the rendering pipeline enables \tool to inspect, and potentially block, all images. Regardless of the image format or how the browser loads it, the raster task decodes the given image into raw pixels, which it then passes to the GPU to display the content on the screen. We run \tool at this precise point to abstract different image formats and loading techniques, while still retaining the opportunity to block an image before the user sees it.

\point{Run multiple instances of \tool in parallel} Running \tool in parallel is a natural design goal because \tool makes all image classification decisions independently based solely on the pixels of each individual image. When designing \tool, we look for opportunities to exploit this natural parallelism to minimize the latency added due to the addition of our ad blocking model.

\subsection{Rendering and \tool: Overview}
We integrate \tool into Blink, the rendering engine for Google Chrome and Brave. From a high level, Blink's primary function is to turn a web page into the appropriate GPU calls~\cite{graphics} to show the user the rendered content.

A web page can be thought of as a collection of HTML, CSS, and JavaScript code, which the browser fetches from the network. The rendering engine parses this code to build the DOM and layout tree, and to issue OpenGL calls via \texttt{Skia}, Google's graphics library~\cite{skia}. 

The layout tree contains the locations of the regions the DOM elements will occupy on the screen. This information together with the DOM element is encoded as a \texttt{display item}.

The browser follows this parsing process by the rasterization process, which takes the display items and turns them into bitmaps. Rasterization issues OpenGL draw calls via the Skia library to draw bitmaps. If the display list items have images in them (a common occurrence), the browser must decode these images before drawing them via Skia.

\tool intercepts the rendering process at this precise point, after the \texttt{Image Decode Task} and during the \texttt{Raster Task}. As the renderer process creates the DOM and decodes and rasterizes all image frames, these are first passed through \tool. \tool blocks the frames that are classified as ads.

\subsection{End-to-End Implementation in Blink}
In our Blink instrumentation, we deal with Skia and Blink classes. Most of the code (forward pass of the CNN) resides at the same directory level as Blink.

Skia uses a set of image decoding operations to turn \texttt{SkImages}, which is the internal class type within Skia that encapsulates images, into bitmaps. \tool reads these bitmaps and classifies their content accordingly. If \tool classifies the bitmap as an ad, we block it by removing its content. Otherwise, \tool lets it pass through to the next layers of the rendering process. In case the content is cleared, we have several options on how to fill up the surrounding white-space. We can either collapse it by propagating the information upwards or display a predefined image (user's spirit animal) in place of the ad.

Figure~\ref{fig:inst} shows an overview of our Blink integration.
Blink class \texttt{BitmapImage} creates an instance of \texttt{DeferredImageDecoder} which in turn instantiates a \texttt{SkImage} object for each encoded image. SkImage creates an instance of \texttt{DecodingImageGenerator} (blink class) which will in turn decode the image using the relevant image decoder from Blink. Note that the image hasn't been decoded yet since chromium practices deferred image decoding.

Finally, \texttt{SkImageGenerator} allocates bitmaps corresponding to the encoded \texttt{SkImage}, and calls \texttt{onGetPixels()} of \texttt{DecodingImageGenerator} to decode the image data using the proper image decoder. This method populates the buffer (pixels) that contain decoded pixels,  which we pass to \tool along with the image height, width, channels information (\texttt{SKImageInfo}) and other image metadata. \tool reads the image, scales it to~\empirical{$224\times 224\times 4$} (default input size expected by SqueezeNet), creates a tensor, and passes it through the CNN. If \tool determines that the buffer contains an ad, it clears the buffer, effectively blocking the image frame.

Rasterization, image decoding, and the rest of the processing happen on a raster thread. Blink rasters on a per tile basis and each tile is like a resource that can be used by the GPU. In a typical scenario there are multiple raster threads each rasterizing different raster tasks in parallel. \tool runs in each of these worker threads after image decoding and during rasterization, which runs the model in parallel.

As opposed to Sentinel~\cite{Sentinel2018} and Ad Highlighter~\cite{adhighlighter} the input to \tool is not the rendered version of web content; \tool takes in the Image pixels directly from the image decoding pipeline. This is important since with \tool we have access to unmodified image buffers and it helps prevent attacks where publishers modify content of the webpage (including iframes) with overlaid masks (using CSS techniques) meant to fool the ad blocker classifier.
\begin{figure}[tb]
\centering
    \includegraphics[width=0.6\columnwidth]{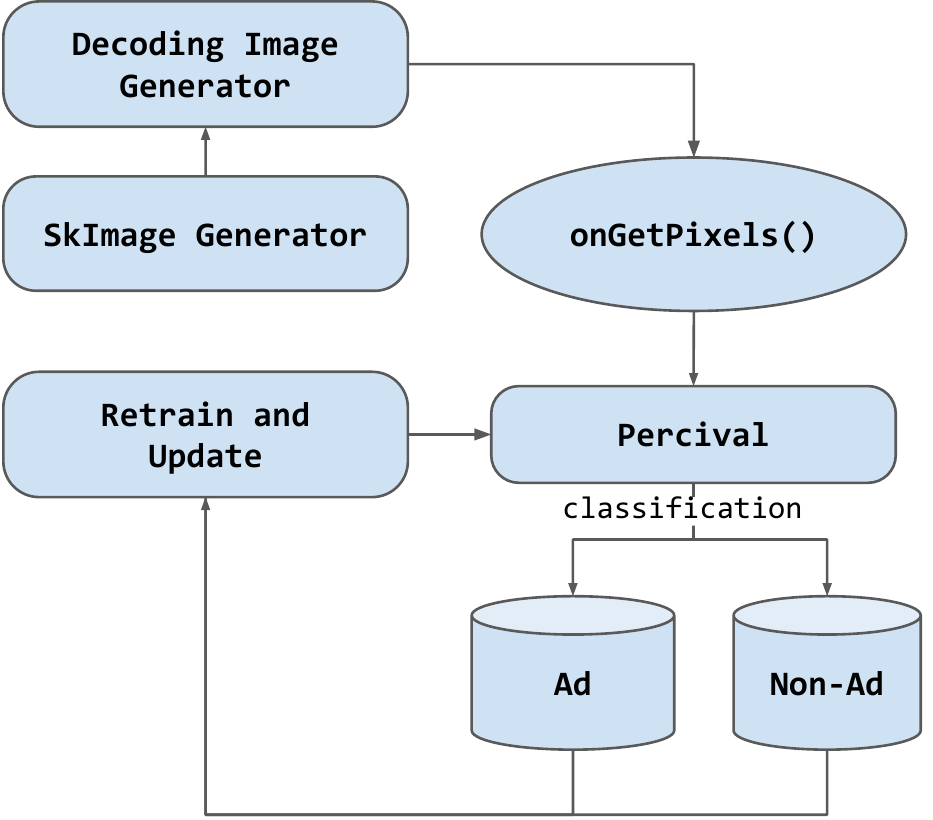}
    \caption{Crawling, labelling and re-training with \tool. Every decoded image frame is passed through \tool and \tool downloads the image frame into the appropriate bucket. 
    }
    \label{fig:crawler}
\end{figure}

\section{Deep Learning Pipeline}
\label{sec:deep-learning}
This section covers the design of \tool's deep neural network and the corresponding training workflow. We first describe the network employed by \tool and the training process. We then describe our data acquisition and labelling techniques.
\subsection{\tool's CNN Architecture}
We cast ad detection as a traditional image classification problem, where we feed images into our model and it classifies them as either being (1) an ad, or (2) not an ad. CNNs are the current standard in the computer vision community for classifying images. 

Because of the prohibitive size and speed of standard CNN based image classifiers, we use a small network, SqueezeNet~\cite{Iandola2016}, as the starting point for our in-browser model.  The SqueezeNet authors show that SqueezeNet achieves comparable accuracy to much larger CNNs, like AlexNet~\cite{Krizhevsky:2012:ICD:2999134.2999257}, and boasts a final model size of~\empirical{4.8~MB}.

SqueezeNet consists of multiple~\emph{fire modules}. A~\emph{fire module} consists of a ``squeeze'' layer, which is a convolution layer with~$1\times 1$ filters and two ``expand'' convolution layers with filter sizes of $1\times 1$ and $3\times 3,$ respectively. Overall, the ''squeeze'' layer reduces the number of input channels to larger convolution filters in the pipeline. 

\begin{figure}[t]
    \centering
    \includegraphics[width=0.65\columnwidth]{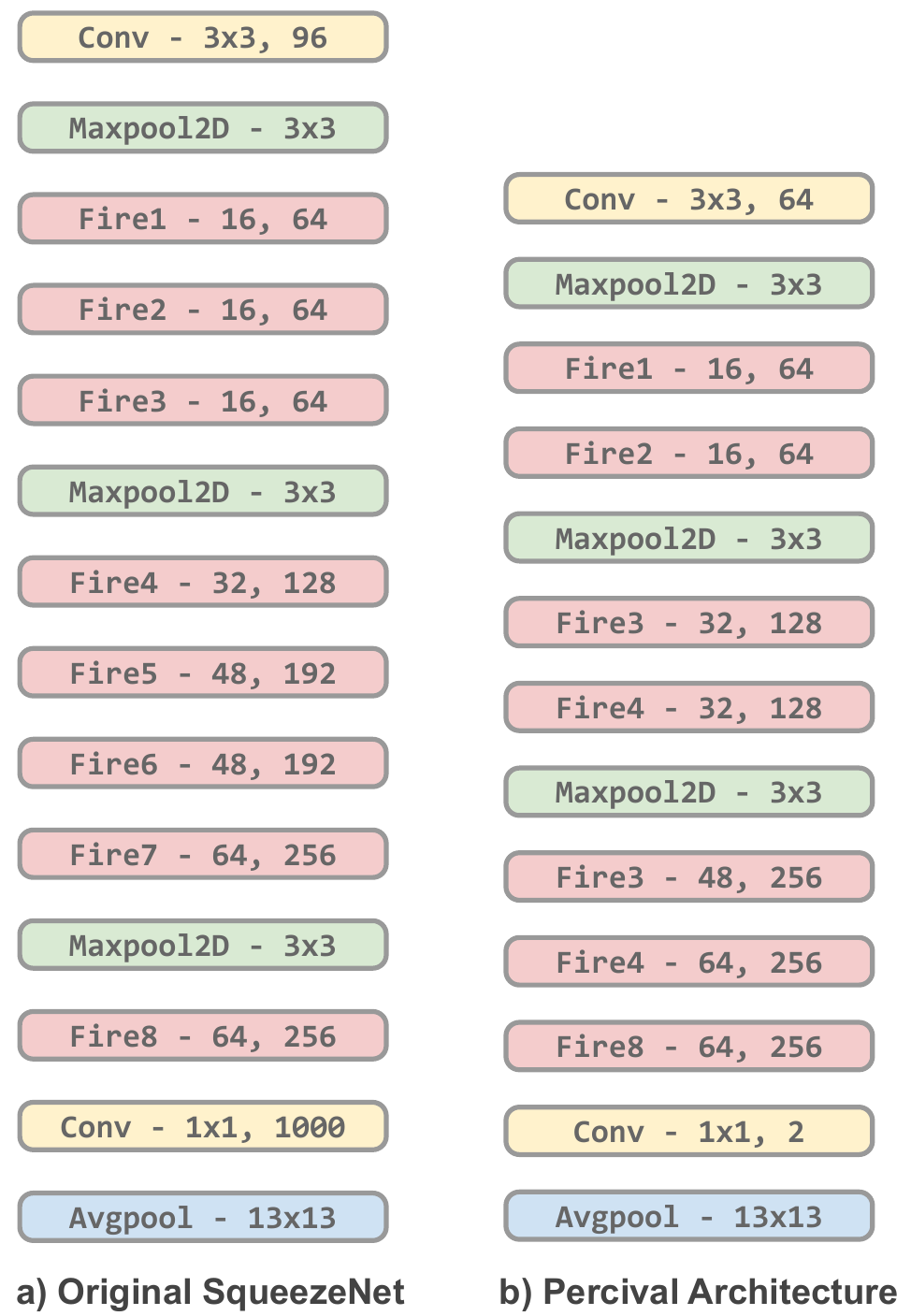}
    \caption{Original SqueezeNet (left) and \tool's fork of SqueezeNet (right). For \texttt{Conv}, \texttt{Maxpool2D}, and \texttt{Avgpool} blocks $a \times b$ represents the dimensions of the filters used. For fire blocks \emph{a, b} represents the number of intermediate and output channels. We remove extraneous blocks as well as downsample the feature maps at regular intervals to reduce the classification time per image.}
    \label{fig:my_cnn}
\end{figure}

A visual summary of \tool's network structure is shown in Figure~\ref{fig:my_cnn}. 
As opposed to the original SqueezeNet, we down-sample the feature maps at regular intervals in the network. This helps reduce the classification time per image. We also perform max-pooling after the first convolution layer and after every two fire modules.



\subsection{Data Acquisition}
We use two systems to collect training image data. First, we use a traditional crawler and traditional ad-blocking rules (EasyList \cite{easylist}) to identify ad images. Second, we use our browser instrumentation from \tool to collect images, improving on some of the issues we encountered with our traditional crawler.

\subsubsection{Crawling with EasyList}
We use a traditional crawler matched with a traditional rule-based ad blocker to identify ad content for our first dataset. In particular, to identify ad elements which could be iframes or complex JavaScript constructs, we use EasyList, which is a set of rules that identify ads based on the URL of the elements, location within the page, origin, class or id tag, and other hand-crafted characteristics known to indicate the presence of ad content.

We built a crawler using Selenium \cite{selenium} for browser automation. We then use the crawler to visit Alexa top-1,000 web sites, waiting for~\empirical{5} seconds on each page, and then randomly selecting~\empirical{3} links and visiting them, while waiting on each page for a period of~\empirical{5} seconds as before. For every visit, the crawler applies every EasyList network, CSS and exception rule.

For every element that matches an EasyList rule, our crawler takes a screenshot of the component, cropped tightly to the coordinates reported by Chromium, and then stores it as an ad sample. We capture non-ad samples by taking screenshots of the elements that do \emph{not} match any of the EasyList rules. Using this approach we, extract~\empirical{22,670} images out of which~\empirical{13,741} are labelled as ads, and~\empirical{8,929} as non-ads. This automatic process was followed by a semi-automated post-processing step, which includes removing duplicate images, as well as manual spot-checking for misclassified images.

Eventually, we identify 2,003 ad images and 7,432 non-ad images. The drop in the number of ad images from 13,741 to 2,003 is due to a lot duplicates and content-less images due to the asynchrony of iframe-loading and the timing of the screenshot. These shortcomings motivated our new crawler. To balance the positive and negative examples in our dataset so the classifier doesn't favor one class over another, we limited the number of non ad and ad images to 2,000.

\subsubsection{Crawling with \tool}
\label{sec:crawling_with_tool}
We found that traditional crawling was good enough to bootstrap the ad classification training process, but it has the fundamental disadvantage that for dynamically-updated elements, the meaningful content is often unavailable at the time of the screenshot, leading to screenshots filled with white-space. 

More concretely, the~\emph{page load} event is not very reliable when it comes to loading iframes. Oftentimes when we take a screenshot of the webpage after the~\emph{page load} event, most of the iframes don't appear in the screenshots. Even if we wait a fixed amount of time before taking the screenshot, iframes constantly keep on refreshing, making it difficult to capture the rendered content within the iframe consistently. 

To handle dynamically-updated data, we use \tool's browser architecture to read all image frames after the browser has decoded them, eliminating the race condition between the browser displaying the content and the screenshot we use to capture the image data. This way we are guaranteed to capture all the iframes that were rendered, independently of the time of rendering or refresh rate.

\point{Instrumentation}
Figure~\ref{fig:crawler} shows how we use \tool's browser instrumentation to capture image data. Each encoded image invokes an instance of \texttt{DecodingImageGenerator} inside Blink, which in turn decodes the image using the relevant image decoder (PNG, GIFs, JPG, etc.). We use the buffer passed to the decoder to store pixels in a bitmap image file, which contains exactly what the render engine sees. Additionally, the browser passes this decoded image to \tool, which determines whether the image contains an ad. This way, every time the browser renders an image, we automatically store it and label it using our initially trained network, resulting in a much cleaner dataset.

\point{Crawling}
To crawl for ad and non-ad images, we run our \tool-based crawler with a browser automation tool called  Puppeteer~\cite{puppeteer}. In each phase, the crawler visits the landing page of each Alexa top-1,000 websites, waits until \texttt{networkidle0} (when there are no more than~0 network connections for at least~500 ms) or~60 seconds. We do this to ensure that we give the ads enough time to load. Then our crawler finds all internal links embedded in the page. Afterwards, it visits~\empirical{20} randomly selected links for each page, while waiting for \texttt{networkidle0} event or~\empirical{60} seconds time out on each request.

In each phase, we crawl between~\empirical{40,000} to~\empirical{60,000} ad images. We then post process the images to remove duplicates, leaving around~\empirical{15-20\%} of  the collected results as useful. We crawl for a total of~\empirical{8} phases, retraining \tool after each stage with the data obtained from the current and all the previous crawls. As before, we cap the number of non-ad images to the amount of ad image to ensure a balanced dataset.

This process was spread-out in time over~\empirical{4} months, repeated every ~\empirical{15} days for a total of~\empirical{8} phases, where each phase took~\empirical{5} days. Our final dataset contains~\empirical{63,000} unique images in total with a balanced split between positive and negative samples.

\begin{figure}[!t]
\centering
\small
\setlength{\tabcolsep}{6pt}
\begin{tabular}{lrrrr}
\toprule
\textbf{Images} &
\textbf{Ads Identified} &
\textbf{Accuracy} &
\textbf{Precision} &
\textbf{Recall} \\
\midrule
6,930 & 3466 & 96.76\% & 97.76\% & 95.72\% \\
\bottomrule
\end{tabular}
\caption{Summary of the results obtained by testing the dataset gathered using EasyList with \tool.}
\label{fig:easylist-eval}
\end{figure}

\section{Evaluation}
\label{sec:eval}
\subsection{Accuracy Against EasyList}
\label{sec:filter-list-comparison}
To evaluate whether \tool can be a viable shield against ads, we conduct a comparison against the most popular crowd-sourced ad blocking list,  EasyList~\cite{easylist}, currently being used by extensions such as Adblock Plus~\cite{adblockplus}, uBlock Origin~\cite{ublock} and Ghostery~\cite{ghostery}.

\begin{figure}[!b]
    \centering
    \small
    \setlength{\tabcolsep}{5pt}
    \begin{tabular}{lrrrrrrrrr}
    \toprule
    \textbf{Ads} &
    \textbf{No-ads}  &
    \textbf{Accuracy} &
    \textbf{FP} &
    \textbf{FN} &
    \textbf{Precision} &
    \textbf{Recall}\\
    \midrule
    354 & 1,830 & 92.0\% & 68 & 106 & 78.4\% & 70.0\% \\
    \bottomrule
\end{tabular}
\caption{Online evaluation of Facebook ads and sponsored content.}
\label{fig:perf_facebook_raw}
\end{figure}

\point{Methodology}
For this experiment, we crawl Alexa top~500 news websites as opposed to Alexa top~1000 websites used in the crawl for training. This is because news websites are an excellent source of advertisements~\cite{news-ads} and the crawl can be completed relatively quickly. Also, Alexa top~500 news websites serves as a test domain different from the train domain we used previously. 

For our comparison we create two data sets: 
First, we apply EasyList rules to select DOM elements that potentially contain ads (IFRAMEs, DIVs, etc.); we then capture screenshots of the contents of these elements. Second, we use resource-blocking rules from EasyList to label all the images of each page according to their resource URL.
After crawling, we manually label the images to identify the false positives resulting in a total of 6,930 images. 
\point{Performance}
On our evaluation dataset, \tool is able to replicate the EasyList rules with accuracy~\empirical{96.76\%}, precision~\empirical{97.76\%} and recall~\empirical{95.72\%} (Figure~\ref{fig:easylist-eval}), illustrating a viable alternative to the manually-curated filter-lists.

\subsection{Blocking Facebook Ads}

\begin{figure}[tb]
\centering
    \small
    \centering
     \includegraphics[width=0.9\columnwidth]{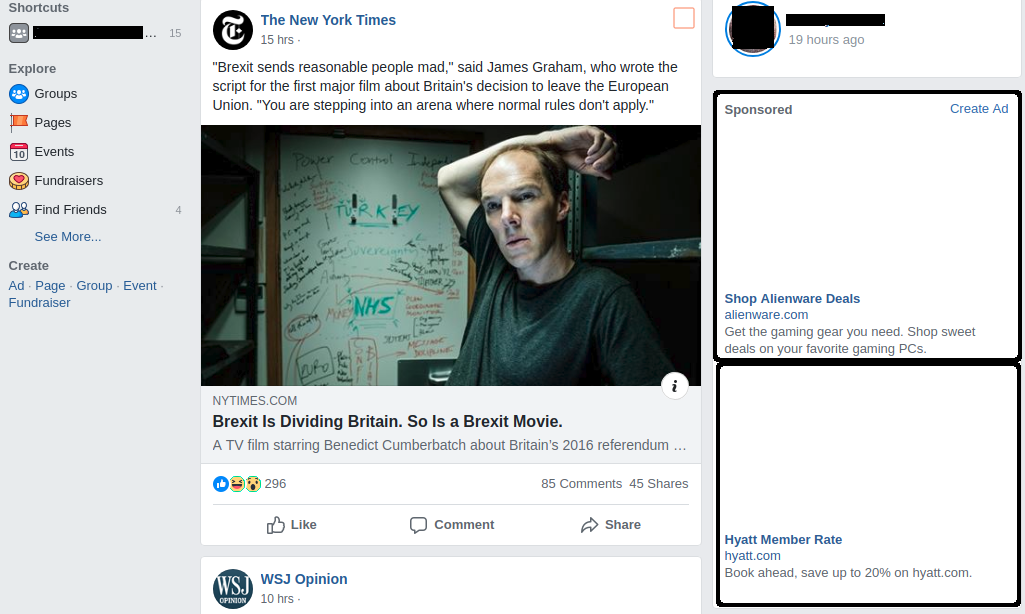}
\caption{The screenshots show one of the author' Facebook home page accessed with \tool. The black rectangles are not part of the original screenshot.}
\label{fig:deepbrave_facebook1}
\end{figure}

\label{sec:facebook}
Facebook obfuscates the ``signatures'' of ad elements (e.g. HTML classes and identifiers) used by filter lists to block ads since its business model depends on serving first-party ads. 
As of now, Facebook does not obfuscate the content of sponsored posts and ads due to the regulations regarding misleading advertising~\cite{abp-no-work, legal}. Even though this requirement favors perceptual ad blockers over traditional ones, a lot of the content on Facebook is user-created which complicates the ability to model ad and non-ad content.

In this section, we assess the accuracy of \tool on blocking Facebook ads and sponsored content. 

\point{Methodology} To evaluate \tool's performance on Facebook, we browse Facebook with \tool for a period of~\empirical{35} days using two non-burner accounts that have been in use for over~\empirical{9} years. Every visit is a typical Facebook browsing session, where we browse through the feed, visit friends' profiles, and different pages of interest. For desktop computers two most popular places to serve ads is the right-side columns and within the feed (labelled sponsored)~\cite{facebookad2}.

For our purposes, we consider content served in these elements as ad content and everything else as non-ad content. A false positive~(FP) is defined as the number of non-ads incorrectly blocked and false negative~(FN) is the number of ads \tool missed to block. For every session, we manually compute these numbers. Figure~\ref{fig:perf_facebook_raw} shows the aggregate numbers from all the browsing sessions undertaken. Figure~\ref{fig:deepbrave_facebook1} shows \tool blocking right-side columns correctly. 

\begin{figure}[h!]
    \centering
    \begin{subfigure}[t]{0.48\columnwidth}
        \centering
        \includegraphics[width=1\textwidth]{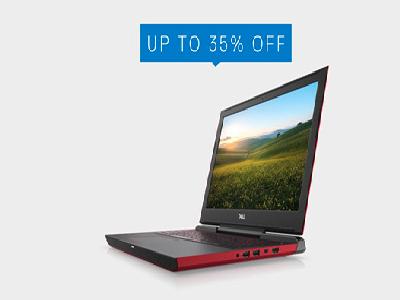}
    \end{subfigure}%
   ~ 
    \begin{subfigure}[t]{0.48\columnwidth}
        \centering
        \includegraphics[width=1\textwidth]{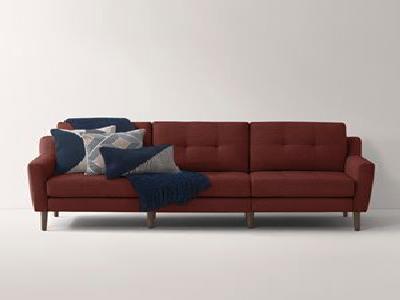}
    \end{subfigure}
    \caption{Examples of false positives and false negatives on Facebook (left) \textbf{False Positive:} This post was created by page owned by Dell Corp. (right) \textbf{False Negative:} This post was part of the sponsored content in the news feed.}
    \label{fig:miss-facebook}
\end{figure}

\point{Results}
Our experiments show that \tool blocks ads on Facebook with a~\empirical{92\%} accuracy and~\empirical{78.4\%} and~\empirical{70.0\%} as precision and recall, respectively. Figure~\ref{fig:perf_facebook_raw} shows the complete results from this experiment.
Even though we achieve the accuracy of~\empirical{92\%}, there is a considerable number of false positives and false negatives, and as such, precision and recall are lower. The classifier always picks out the ads in the right-columns but struggles with the ads embedded in the feed. This is the source of majority of the false negatives. False positives come from high ``ad intent'' user-created content, as well as content created by brand or product pages on Facebook (Figure~\ref{fig:miss-facebook}).

\point{Discussion: False Positives and False Negatives} To put Figure~\ref{fig:perf_facebook_raw} into perspective since it might appear to have an alarming number of false positives and false negatives, it is worthwhile to consider an average scenario. If each facebook visit on average consists of browsing through 100 images, then by our experiments, a user will find roughly 16 ad images and 84 non-ad images, out of which \tool will block 11 to 12 ad images on average while also blocking 3 to 4 non-ad images. This is shown in Figure~\ref{fig:perf_facebook_average}. 

In addition to the above mentioned experiments which evaluate the out of box results of using \tool, we trained a version of \tool on a particular user's ad images. The model achieved higher precision and recall of 97.25\%, 88.05\% respectively.

\begin{figure}[t!]
    \centering
    \small
    \setlength{\tabcolsep}{3pt}
    \begin{tabular}{lrrrrr}
    \toprule
    \textbf{Language} &
    \textbf{\# crawled} &
    \textbf{\# Ads} &
    \textbf{Accuracy}  &
    \textbf{Precision} &
    \textbf{Recall}\\
    \midrule
    Arabic & 5008 & 2747 & 81.3\% & 83.3\% & 82.5\% \\
    Spanish & 2539& 309 & 95.1\% & 76.8\% & 88.9\% \\
    French & 2414 &366 &93.9\% & 77.6\% & 90.4\% \\
    Korean & 4296 & 506 & 76.9\% & 54.0\% & 92.0\% \\
    Chinese & 2094 & 527 & 80.4\% & 74.2\% & 71.5\% \\
    \bottomrule
\end{tabular}
\caption{Accuracy of \tool on ads in non-English languages. The second column represents the number of images we crawled, while the third column is the number of images that were identified as ads by a native speaker. The remaining columns indicate how well \tool is able to reproduce these labels.}
\label{fig:international}
\end{figure}

\begin{figure}[t!]
    \centering
    \small
    \setlength{\tabcolsep}{9pt}
    \begin{tabular}{lrrrrrrr}
    \toprule
    \textbf{Images} &
    \textbf{Ads} &
    \textbf{No-ads}  &
    \textbf{FP} &
    \textbf{FN} &\\
    \midrule
    100 & 16 & 84 & 3-4 & 4-5\\
    \bottomrule
\end{tabular}
\caption{Average reporting of evaluation of Facebook ads and sponsored content per visit. We assume each Facebook visit consists of browsing through 100 total images.}
\label{fig:perf_facebook_average}
\end{figure}

\begin{figure}[b!]
    \includegraphics[width=\columnwidth,height=0.5\columnwidth]{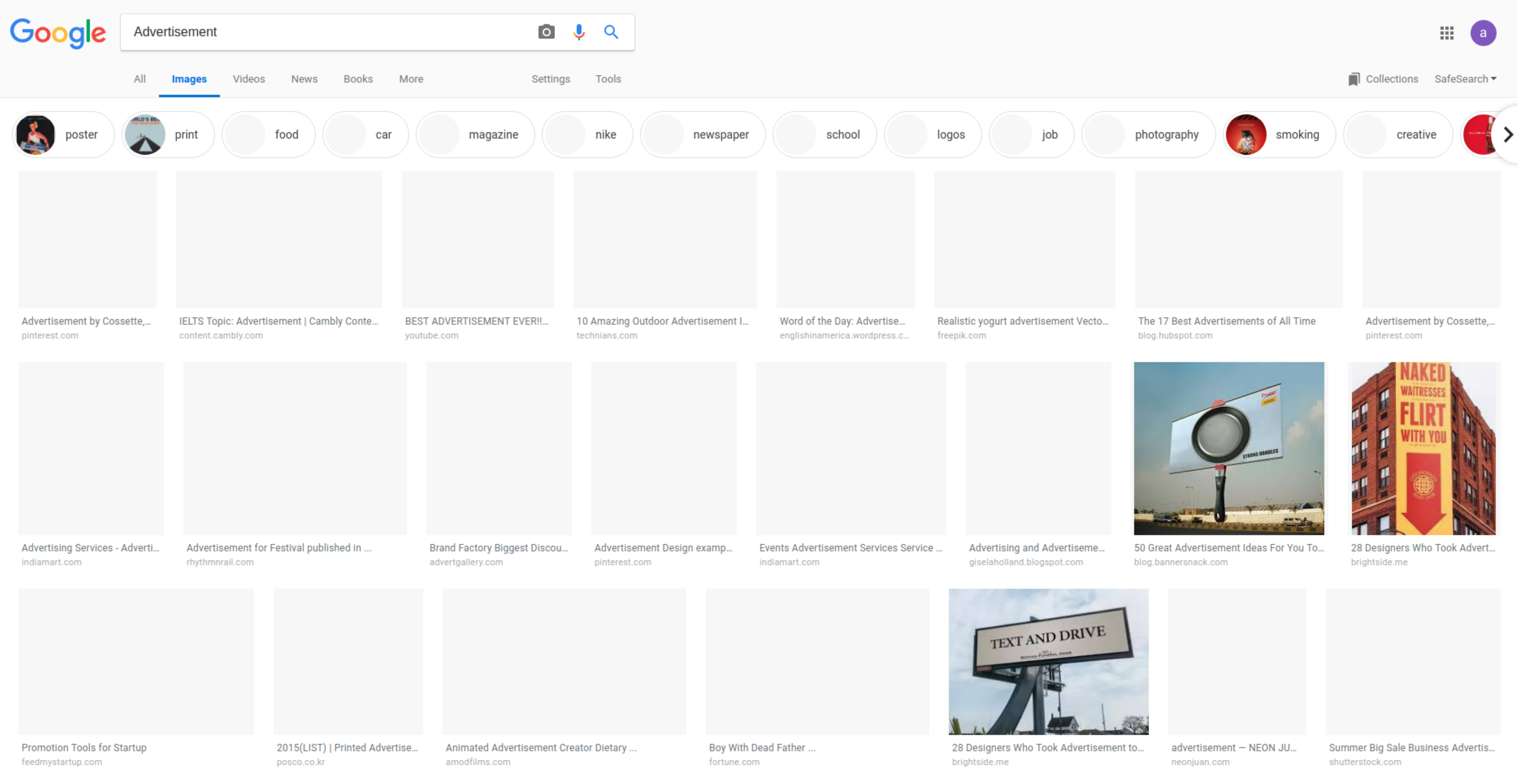}
    \caption{Search results from searching for ``Advertisement'' on Google images, using \tool.}
    \label{fig:deepbrave_Advertisement}
\end{figure}

\subsection{Blocking Google Image Search Results}
\label{sec:google_images}
To improve our understanding of the misclassifications of \tool, we used Google Images as a way to fetch images from distributions that have high or low ad intent. For example, we fetched results with the query ``Advertisement'' and used \tool to classify and block images. As we can see in Figure~\ref{fig:deepbrave_Advertisement}, out of the top~23 images,~20 of them were successfully blocked. Additionally, we tested with examples of low ad intent distribution we used the query ``Obama''). We also searched for other keywords, such as ``Coffee'', ``Detergent'', etc. The detailed results are presented in Figure~\ref{fig:deepbrave_screenshot}. As shown, \tool can identify a significant percentage of images on a highly ad-biased content.


\begin{figure}[h!]
\centering
\setlength{\tabcolsep}{8pt}
\footnotesize
\begin{tabular}{lrrrr}
    \toprule
     \bf Query & \bf \# blocked & \bf \# rendered & \bf FP & \bf FN \\\midrule
     Obama  & 12 & 88 & 12 & 0 \\
     Advertisement& 96 & 4 & 0 & 4  \\
     Coffee & 23 & 77   & - & -\\
     Detergent & 85 & 15 & 10 & 6\\
     iPhone & 76 & 24 & 23 & 1\\
    \bottomrule
\end{tabular}
\caption{\tool blocking image search results. For each search we only consider the first \empirical{100} images returned (``-'' represents cases where we were not able to determine whether the content served is ad or non-ad). }
\label{fig:deepbrave_screenshot}
\end{figure}

\subsection{Language-Agnostic Detection}
\label{sec:language-agnostic}

We test \tool against images with language content different than the one we trained on. In particular, we source a data set of images in Arabic, Chinese, French, Korean and Spanish.
\point{Crawling}
To crawl for ad and non-ad images, we use ExpressVPN~\cite{expressvpn} to VPN into major world cities where the above mentioned languages are spoken. For instance, to crawl Korean ads, we VPN into two locations in Seoul. We then manually visit top 10 websites as mentioned in SimilarWeb~\cite{similarweb} list. We engage with the ad-networks by clicking on ads, as well as closing the ads (icon at the top right corner of the ad) and then choosing random responses like content not relevant or ad seen multiple times. This is done to ensure we are served ads from the language of the region.

We then run \tool-based crawler with the browser automation tool Puppeteer~\cite{puppeteer}. Our crawler visits the landing page of each top 50 SimilarWeb websites for the given region, waits until \texttt{networkidle0} (when there are no more than~0 network connections for at least~500 ms) or~60 seconds. Then our crawler finds all internal links embedded in the page. Afterwards, it visits~\empirical{10} randomly selected links for each page, while waiting for \texttt{networkidle0} event or~\empirical{60} seconds time out on each request.
As opposed to Section~\ref{sec:crawling_with_tool}, we download every image frame to a single bucket.

\begin{figure}[b!]
    \centering
    \begin{subfigure}[t]{0.48\columnwidth}
        \centering
        \includegraphics[width=1\textwidth]{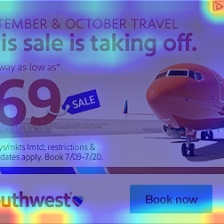}
        \caption{Ad image: Layer 9}
    \end{subfigure}%
   ~ 
    \begin{subfigure}[t]{0.48\columnwidth}
        \centering
        \includegraphics[width=1\textwidth]{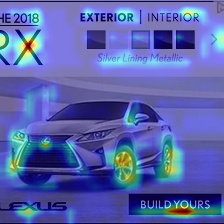}
        \caption{Ad image: Layer 5}
    \end{subfigure}
    \caption{Salience map of the network on a sample ad images. Each image corresponds to the output of Grad-CAM~\cite{SelvarajuDVCPB16} for the layer in question.}
    \label{ref:saliency}
\end{figure}

\begin{figure}[tbp]
\centering
\centering
\includegraphics[width=1.0\columnwidth]{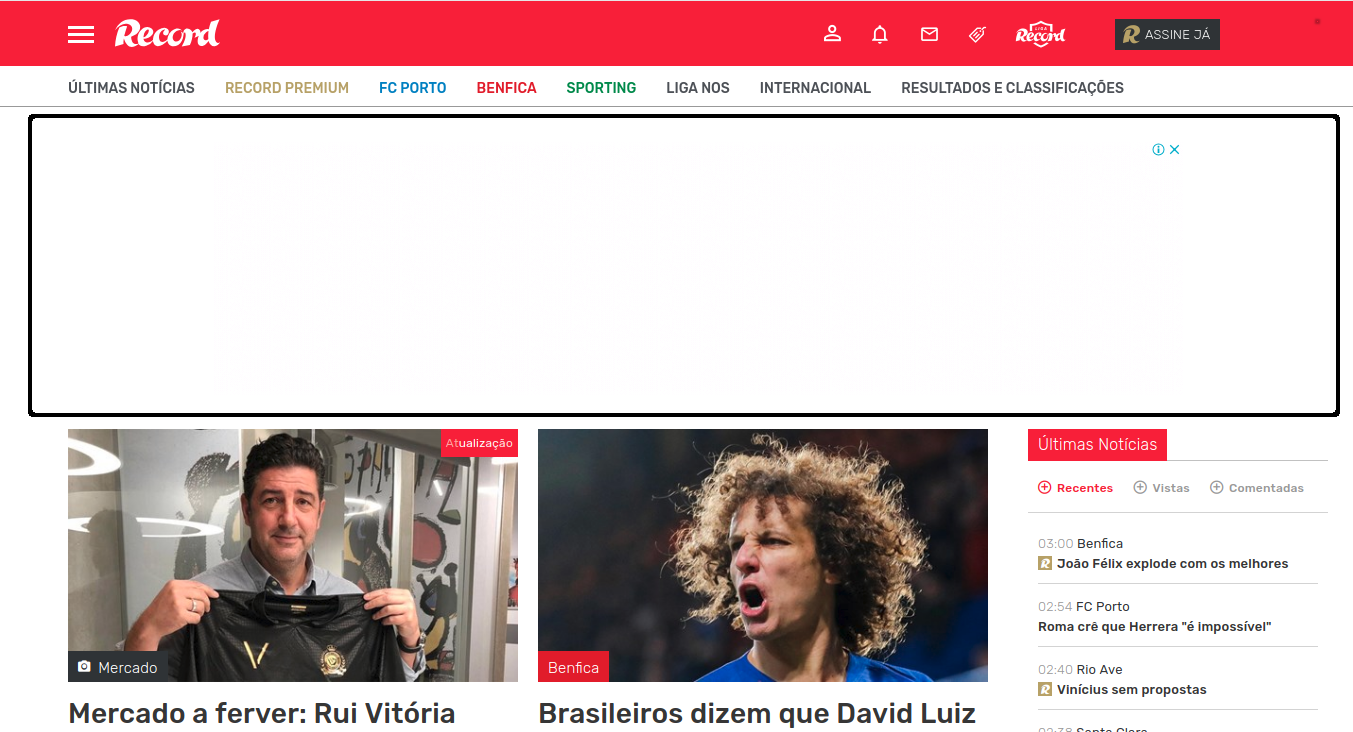}
\caption{\tool results on \url{record.pt} (Portuguese language website).}
\label{fig:deepbrave_korea}
\end{figure}

\point{Labeling}
For each language, we crawl~\empirical{2,000--6,000} images. We then hire a native speaker of the language under consideration and have them label the data crawled for that language. Afterwards, we test \tool with this labeled dataset to determine how accurately can \tool reproduce these human annotated labels. Figure~\ref{fig:international} shows the detailed results from all languages we test on. Figure~\ref{fig:deepbrave_korea} shows a screen shot of a Portuguese website rendered with \tool.
\point{Results}
Our experiments show that \tool can generalize to different languages with high accuracy (\empirical{81.3\%} for Portuguese,~\empirical{95.1\%} for Spanish,~\empirical{93.9\%} for French) and moderately high precision and recall (\empirical{83.3\%, 82.5\%} for Arabic,~\empirical{76.8\%, 88.9\%} for Spanish,~\empirical{77.6\%, 90.4\%} for French). This illustrates the out-of-the box benefit of using \tool for languages that have much lower coverage of EasyList rules, compared to the English ones. The model does not perform as well on Korean and Chinese datasets.

\subsection{Salience Map of the CNN}
\label{sec:salience}
To visualize which segments of the image are influencing the classification decision, we used Grad-CAM~\cite{SelvarajuDVCPB16} network salience mapping which allow us to highlight the important regions in the image that caused the prediction. As we can see in Figure~\ref{ref:saliency}, our network is focusing on ad visual cues (\textit{AdChoice} logo), when this is present (case (a)), also it follows the outlines of text (signifying existence of text between white space) or identifies features of the object of interest (wheels of a car). 

\subsection{Runtime Performance Evaluation}
\label{sec:runtime-perf}

We next evaluate the impact of \tool-based blocking on the browser performance. This latency is a function to the number and complexity of the images on the page and the time the classifier takes to classify each of them. We measure the rendering time impact when we classify each image \emph{synchronously}.

To evaluate the performance of our system, we used top~5,000 URLs from Alexa to test against Chromium compiled on Ubuntu Linux~16.04, with and without \tool activated. We also tested \tool in Brave, a privacy-oriented Chromium-based browser, which blocks ads using block lists by default. For each experiment we measured \texttt{render} time which is defined as the difference between \texttt{domComplete} and \texttt{domLoading} events timestamps. We conducted the evaluations sequentially on the same Amazon m5.large EC2 instance to avoid interference with other processes and make the comparison fair. Also, all the experiments were using \texttt{xvfb} for rendering, an in-memory display server which allowed us to run the tests without a display.

\begin{figure}[t!]
    \centering
    \includegraphics[width=.8\columnwidth]{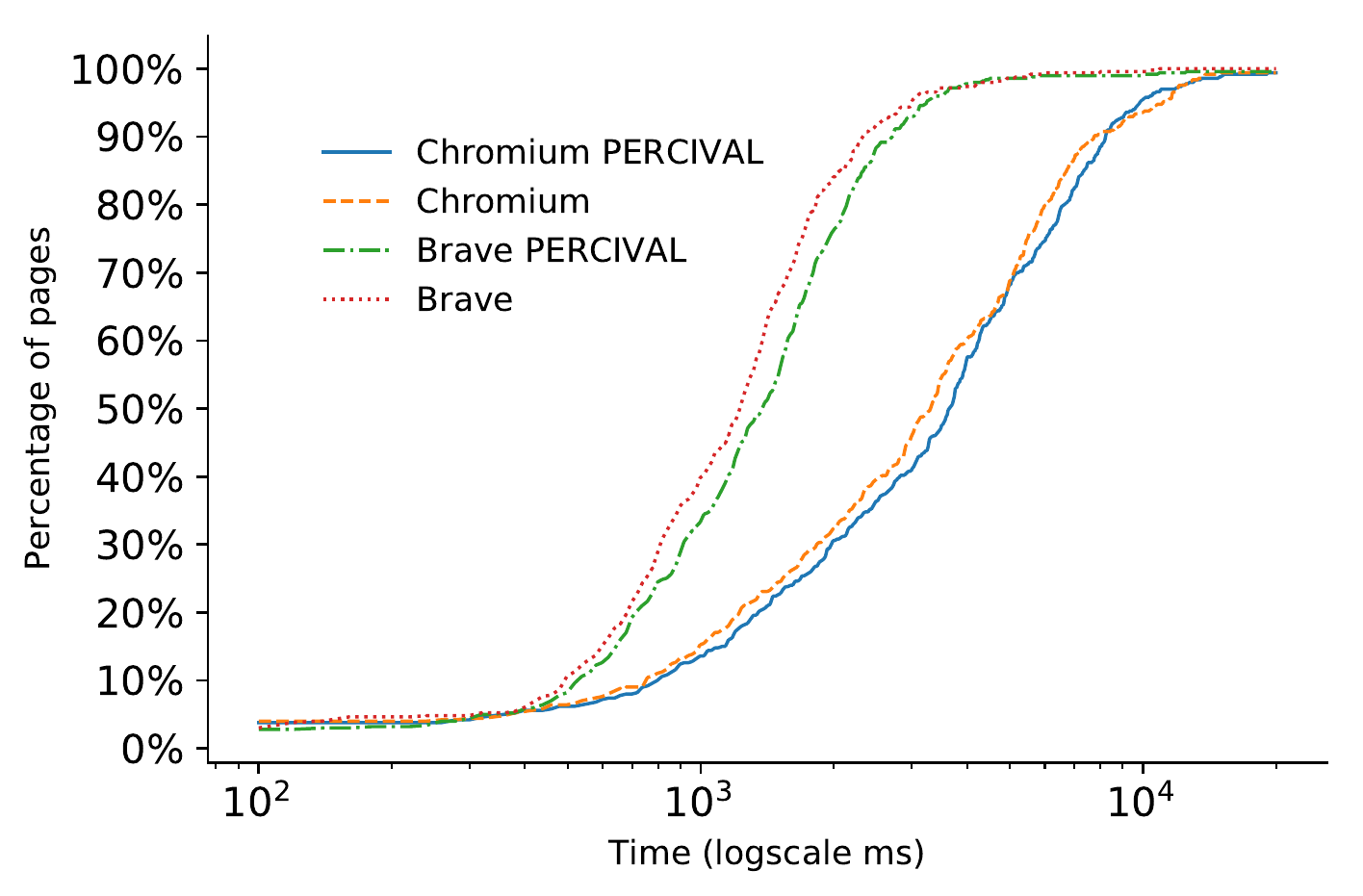}
    \caption{\texttt{Render} time evaluation in Chromium and Brave browser.}
    \label{fig:perf_cdf}
\end{figure}

In our evaluation we show an increase of~\emperical{178.23}ms of median render time when running \tool in the rendering critical path of Chromium and~\emperical{281.85}ms when running inside Brave browser with ad blocker and shields on. Figures~\ref{fig:perf_cdf} and~\ref{fig:perf_table} summarize the results.

To capture rendering and perceptual impact better, we create a micro-benchmark with \texttt{firstMeaningfulPaint} to illustrate overhead. In our new experiment, we construct a static html page containing 100 images. We then measure \texttt{firstMeaningfulPaint} with Percival classifying images synchronously and asynchronously. In synchronous classification, \tool adds 120ms to Chrome and 140ms to Brave. In asynchronous classification, \tool adds 6ms to Chrome and 3ms to Brave. Although asynchronous classification nearly eliminates overhead, it opens up the possibility of showing an image to the user that we later remove after flagging it as an ad because the rasterization of the image runs in parallel with classification in this mode of operation.

To determine why \tool with Brave is slower than Chromium. We trace events inside the decoding process using \texttt{firstMeaningfulPaint} and confirm there is no significant deviation between the two browsers. The variance observed initially is due to the additional layers in place like Brave's ad blocking shields.

\begin{figure}[b!]
  \footnotesize
  \centering
  \setlength{\tabcolsep}{6pt}
  \begin{tabular}{llrr}
    \toprule
    \bf Baseline & \bf Treatment  & \bf Overhead (\%) & \bf (ms) \\ 
    \midrule
    Chromium & Chromium + \tool & 4.55 & 178.23 \\
    Brave & Brave + \tool       &  19.07 & 281.85 \\
    \bottomrule
  \end{tabular}
  \caption{Performance evaluation of \tool on \texttt{Render} metric.}
  \label{fig:perf_table}
\end{figure}

\subsection{Comparison With Other Deep Learning Based Ad Blockers}
\label{sec:comparison-sentinel}

Recently, researchers evaluated the accuracy of three deep-learning based perceptual ad blockers including \tool~\cite{Tramer2018}. They used real website data from Alexa top~10 news websites to collect data which is later manually labelled. In this evaluation, \tool outperformed models 150 times bigger than \tool in terms of recall. We show their results in Figure~\ref{fig:tramer}.

\subsection{Adversarial Attacks against \tool}
\label{sec:adv-eval-percep}
In recent work by Tramèr~\etal~\cite{Tramer2018}, they show how the implementation of some state-of-the-art perceptual ad blockers, including \tool, is vulnerable to attacks. However, only one of the attacks where they used \tool's model to create adversarial ad images affects \tool due to our design decision to run \tool client-side thereby giving attackers white box access to the model.

To address the concern that adversaries have white box access to the model, we argue that \tool is extremely light-weight and can be re-trained and updated very quickly. Our model currently takes 9 minutes (7 epochs) to fine-tune the weights of the network on an NVIDIA V 100 GPU, meaning that we can generate new models very quickly. \tool is 1.7MB which is almost half the average web page in 2018~\cite{page-size} making frequent downloads easier.

To demonstrate, re-training and model update as an effective defense against the adversarial samples, we trained a MobilenetV2~\cite{mobilenetv2} with our current dataset. It took 9 minutes of fine-tuning to get to our baseline accuracy. The updated model correctly classified all the adversarial samples generated for \tool by Tramer~\etal~\cite{Tramer2018} suggesting that none of the samples transferred to this model. It should be noted that, we did not add any more data to our dataset.

Additionally, to improve the robustness of the models against adversarial attacks one could employ techniques like \emph{min-max} (robust) optimization~\cite{madry2017towards} ,where the classification loss is minimized while maximizing the acceptable perturbation one can apply to the image, or \emph{randomized smoothing} \cite{li2019certified,pmlrv97cohen19c,lecuyer2019certified} where provable (or certified) robust accuracy can be afforded. Such techniques have shown promising results in training robust models and are currently under active research \cite{zhang2019you,shafahi2019adversarial}.

While we do accept that given sufficient time and machine learning expertise, it may be possible to create adversarial samples that generalize across different models but it in effect makes evasion more expensive. If we can update the model frequently, adversaries will have to play catch-up every time.

The authors in~\cite{Tramer2018} also claim that one adversarial sample influences \tool to block another benign non-ad image. This, however, is not true; the authors claim to use two benign images, one of which is not benign and other is contentless white-space image. \tool blocks these images. If these are replaced with stock non-ad images, \tool correctly renders both, meaning that \tool makes each decision independently and is not vulnerable to hijacking as is claimed in the paper. 


\begin{figure}[tb]
\centering
\small
\setlength{\tabcolsep}{4pt}
\begin{tabular}{lrrrr}
\toprule
\textbf{Model} &
\textbf{Size} &
\textbf{FP} &
\textbf{FN} & \\
\midrule
Sentinel\cite{sentinel} Clone & 256 MB & 0/20 & 5/29 \\
ResNet~\cite{Hussain2017} & 242 MB & 0/20 & 21/39 \\
\tool & 1.76 MB & 2/7 & 3/33 \\
\bottomrule
\end{tabular}
\caption{Tramer~\etal's~\cite{Tramer2018} evaluation of various deep learning based perceptual ad blockers. The difference in the number of images used for evaluation stem from the kind of images the ad blocker is expecting. For instance, the in-browser \tool model ignores all image frames smaller than 100x100 for performance reasons.}
\label{fig:tramer}
\end{figure}

\section{Limitations}
\label{sec:discussion}

\point{Dangling Text}
By testing \tool integrated into Chromium, we noticed the following limitations. Many ads consist of multiple elements, which contain images and text information layered together. \tool is positioned in the rendering engine, and therefore it has access to one image at a time. This leads to situations where we effectively block the image, but the text is left dangling. Although this is rare, we can mitigate this by retraining the model with ad image frames containing just the text. Alternatively, a non-machine learning solution would be to memorize the DOM element that contains the blocked image and filter it out on consecutive page visitations. Although this might provide an unsatisfying experience to the user, we argue that it is of the benefit of the user to eventually have a good ad blocking experience, even if this is happening on a second page visit.

\point{Small Images} Currently, images that are below $100 \times 100$ size bypass \tool to reduce the processing time. This is a limitation which can be alleviated by deferring the classification and blocking of small images to a different thread, effectively blocking \emph{asynchronously}. That way we make sure that we don't regress the performance significantly, while we make sure that consecutive requests will continue blocking small ads.



\section{Related Work}
\label{sec:related}



\point{Filter lists} 
Popular ad blockers like, Adblock Plus~\cite{adblockplus}, uBlock Origin~\cite{ublock}, and Ghostery~\cite{ghostery} are using a set of rules, called filter-list, to block resources that match a predefined crowd-sourced list of regular expressions (from lists like EasyList and EasyPrivacy). On top of that, CSS rules are applied, to prevent DOM elements that are potential containers of ads. These filter-lists are crowd-sourced and updated frequently to adjust on the non-stationary nature of the online ads ~\cite{Storey}. For example, EasyList, the most popular filter-list, has a history of~9 years and contains more than~\empirical{60.000} rules~\cite{vastel2018filters}. However, filter-list based solutions enable a continuous cat-and-mouse game: their maintenance cannot scale efficiently, as they depend on the human-annotator and they do not generalize to ``unseen'' examples.


\point{Perceptual Ad Blocking}
Perceptual ad blocking is the idea of blocking ads based solely on their appearance; an example ad, highlighting some of the typical components.
Storey~\etal~\cite{Storey} uses the rendered image content to identify ads. More specifically, they use OCR and fuzzy image search techniques to identify \textit{visual cues} such as ad disclosure markers or sponsored content links. Unlike \tool, this work  assumes that the ad provider is complying with the legislation and is using visual cues like \emph{AdChoices}. 

Sentinel system~\cite{Sentinel2018} proposes a solution based on convolutional neural networks~(CNNs) to identify Facebook ads. This work is closer to our proposal; however, their model is not deployable in mobile devices or desktop computers because of its large size (>200MB). 

Also, we would like to mention the work of~\cite{Ye2018,Hussain2017,Ahuja2018}, where they use deep neural networks to identify the represented signifiers in the Ad images. This is a promising direction in semantic and perceptual ad blocking.



\point{Adversarial attacks}
In computer-vision, researchers have demonstrated attacks that can cause prediction errors by near-imperceptible perturbations of the input image. This poses risks in a wide range of applications on which computer vision is a critical component (e.g. autonomous cars, surveillance systems)~\cite{Papernot2016a,Papernot2016b,Papernot2016c}. Similar attacks have been demonstrated in speech to text~\cite{Carlini2018}, malware detection~\cite{grosse2017adversarial} and reinforcement-learning~\cite{huang2017adversarial}. To defend from adversarial attacks, a portfolio of techniques has been proposed~\cite{madry2017towards,kurakin2018ensemble,kolter2017provable, kannan2018adversarial,kurakin2016adversarial,chen-carlini}, whether these solve this open research problem, remains to be seen.
\section{Conclusion}
\label{sec:conclusions}



With \tool, we illustrate that it is possible to devise models that block ads, while rendering images inside the browser. Our implementation inside Chromium shows a rendering time overhead of~\performancerel, demonstrating the feasibility of deploying deep neural networks inside the critical path of the rendering engine of a browser. We show that our perceptual ad blocking model can replicate EasyList rules with an accuracy of~\easylistaccuracy, making \tool  a viable and complementary ad blocking layer. Finally, we demonstrate off the shelf language-agnostic detection due to the fact that our models do not depend on textual information. Finally, we show that \tool is a compelling blocking mechanism for first-party Facebook sponsored content, for which traditional filter based solutions are less effective.


\bibliographystyle{plain}
\bibliography{perceptual_and_attacks}

\end{document}